# Tuning the Band Gap in Silicene by Oxidation

*Yi Du[1*], Jincheng Zhuang[1], Hongsheng Liu[3], Xun Xu[1], Stefan Eilers[1], Kehui Wu[2], Peng Cheng[2], Jijun Zhao[3], Xiaodong Pi[4], Khay Wai See[1], Germanas Peleckis[1], Xiaolin Wang[1] and Shi Xue Dou[1]*

[1] Institute for Superconducting and Electronic Materials (ISEM), University of Wollongong, Wollongong, NSW 2525, Australia

[2] Institute of Physics, Chinese Academy of Science, Haidian District, Beijing 100080, China

[3] Key Laboratory of Materials Modification by Laser, Ion and Electron Beams (Dalian University of Technology), Ministry of Education, Dalian 116024, China

[4] State Key Laboratory of Silicon Materials and Department of Materials Science and Engineering, Zhejiang University, Hangzhou 310027, China,

**ABSTRACT**

Silicene monolayers grown on Ag(111) surfaces demonstrate a band gap that is tunable by oxygen adatoms from semimetallic to semiconducting type. By using low-temperature scanning tunneling microscopy, it is found that the adsorption configurations and amounts of oxygen adatoms on the silicene surface are critical for band-gap engineering, which is dominated by different buckled structures in $\sqrt{13}\times\sqrt{13}$, $4\times4$, and $2\sqrt{3}\times2\sqrt{3}$ silicene layers. The Si-O-Si bonds are the most energy-favored species formed on √13×√13, 4×4, and 2√3×2√3 structures under oxidation, which is verified by *in-situ* Raman spectroscopy as well as first-principles calculations. The silicene monolayers retain their structures when fully covered by oxygen adatoms. Our work demonstrates the feasibility of tuning the band gap of silicene with oxygen adatoms, which, in turn, expands the base of available two-dimensional electronic materials for devices with properties that is hardly achieved with graphene oxide.

Silicene, a new allotrope of silicon in a two-dimensional honeycomb structure, has attracted intensive research interest due to its novel physical and chemical properties.[1-10] Theoretically, strong spin-orbit coupling is predicted in silicene, which may allow a spin-orbit band gap of 1.55 meV at the Dirac point and results in a detectable quantum spin Hall effect (QSHE).[9-13] The electronic π- and π*- bands derived from the Si $p_z$ orbital disperse linearly to cross at the Fermi level ($E_F$), leading to massless Dirac fermion behaviour of electrons.[12-13] Electrons in silicene thus have a large Fermi velocity, which has been recently verified by pronounced quasiparticle interference patterns observed by scanning tunneling microscopy (STM).[12] In addition, excellent scalability and compatibility with current silicon-based nanotechnology make silicene a promising candidate for the design of novel electronic components and interconnects at the nanometer scale.[14]

Despite its remarkable properties, the intrinsic zero band gap in silicene hinders its applications in electronic devices which require controllable conductivities through logic gates. Therefore, tuning the band gap in silicene is highly desirable. Conventionally, chemical doping, selective functionalization, and the introduction of defects are regarded as effective approaches to modulate band structures in two-dimensional (2D) zero-gap materials, as demonstrated in graphene. These approaches can only be associated with edges or surface defects, however, because the carbon atoms located within the plane are relatively chemically inert due to the $sp^2$ hybridization of carbon, while those located at the edges or at defects are more reactive, which limits the functionality of graphene.[15] Unlike carbon atoms in graphene, silicon atoms tend to adopt $sp^3$ hybridization over $sp^2$ in silicene, which makes it highly chemically active on the surface and allows its electronic states to be easily tuned by chemical functionalization.[16,17] Oxygen is one of the feasible species for chemical functionalization, as oxygen atoms are capable of breaking the symmetry so as to alter the electronic structure, such as with the energy-gap opening at $E_F$ in 2D materials.[15] Controllable oxidation is therefore expected to be of significance for modulating electronic states in silicene. It not only provides an opportunity for making various electronic structures in silicene, but also offers the possibility of exploring so-called "silicene oxide" which can be readily used in silicene-based electronic devices, such as in gate oxides. Moreover, oxidation of silicene layers is expected to be one of the major steps towards effective introduction of oxygenated functional groups into the Si network, which will further promote silicene functionalizations for various application purposes. Nevertheless, the high chemical reactivity of silicene prevents controllable oxidation by conventional chemical routes, such as the solvent-casting method, thus obstructing progress in such research.

In this work, we report a study of band-gap tuning in different silicene buckling structures by controllable oxidation processes, using scanning tunneling microscopy and *in-situ* Raman spectroscopy. The correlation between buckled silicene structures and oxygen adatoms is identified with the aid of density functional theory (DFT) calculations. We show that the detailed bonding configurations of oxygen adatoms on the silicene surface rely on the buckling structures. The oxygen adatoms can effectively tune the band gap, which results in gap opening. It is found that silicene possibly retains its honeycomb structure even when its surface is fully covered by oxygen adatoms. In addition, the surface of silicene exhibits much higher chemical reactivity as compared to the edge, which is very different from the case of graphene.[15]

**RESULTS AND DISCUSSION**

Epitaxial silicene monolayers exhibit different reconstructions on the Ag(111) surface, which can be controlled by the substrate temperature during deposition. Figure 1 shows single-layer silicene in three typical structures, namely, $\sqrt{13}\times\sqrt{13}$, 4×4, and $2\sqrt{3}\times2\sqrt{3}$, grown on Ag(111) surfaces at different substrate temperatures. Because of the similar formation energies, √13×√13 and 4×4 structures always coexist when the substrate temperature is between 450 K and 520 K during deposition, as shown in Figure 1(a). When the substrate temperature is further increased to 550 K, the pure 2√3×2√3 silicene structure can be attained [Figure 1(b)]. High-resolution STM images for each structure are shown in Figure 1(c-e). All silicene structures demonstrate distinctive buckled forms that can be distinguished by contrast in the images. In this work, the topmost Si atoms in a buckled structure are defined as the "top-layer" (TL) and the other atomic layers are defined as "bottom-layer" (BL). Unlike the $sp^2$ hybridization for carbon in graphene, silicon atoms exhibit an energetically stable $sp^3$ hybridization, which is responsible for these low-buckled structures.[26] On the Ag(111) surface, various buckling structures of silicene, which exist due to variations in mismatch and interaction between silicene superstructures and the substrate, are shown in Figure 1(f). The metal passivation effect induced by the Ag(111) surface also influences the buckling of silicene on the BL Si side because of hybridization of the $p_z$ electrons of BL Si atoms with the 4$d$ electrons of Ag(111). The height of buckling in these three structures therefore varies from 0.86 Å for the 4×4 structure to 1.40 Å for the √13×√13 structure.[19] In the √13×√13 structure, only one TL Si atom out of the fourteen Si atoms per unit cell resides on the top site in the buckled structure. In contrast, six out of eighteen Si atoms per unit cell are topmost Si in the 4×4 structure, which results in a "flower-like" pattern in STM images. In the 2√3×2√3 phase, Si atoms are in a "three-fold" or "bridge" position on the Ag(111) surface; hence, there are two topmost atoms out of fourteen Si atoms per unit cell. The distances between nearest neighbouring TL Si atoms are 5.46 Å, 2.51 Å, and 3.67 Å for the √13×√13, 4×4, and 2√3×2√3 structures, respectively. TL Si atoms are expected to be highly active in epitaxial silicene on Ag(111), since they are unsaturated atoms, which might be due to the dehybridization effect.[16] BL Si atoms, in contrast, are relatively more stable, as they are passivated by the free electrons from the substrate. The surface chemical reactivity of silicene is, therefore, dominated by the buckled structures because the numbers of TL Si atoms in these three structures are different, *e.g.* the $\sqrt{13}\times\sqrt{13}$ structure has fewer TL Si atoms than the 4×4 and $2\sqrt{3}\times2\sqrt{3}$ structures.

Figure 2(a-c) presents typical STM images of silicene layers in $\sqrt{13}\times\sqrt{13}$, 4×4, and $2\sqrt{3}\times2\sqrt{3}$ structures that were exposed to 10 Langmuir (L) $O_2$. The marked isolated protrusions in these STM images are clearly different from the clean silicene surface in Figure 1. The protrusions are higher than for TL Si atoms. The interpretation of these protrusions is obtained from the scanning tunnelling spectroscopy (STS) maps of $d$I/$d$V, which is determined by the local density of state (LDOS), and *in-situ* Raman spectroscopy, which reflects vibrational modes of chemical bonds. By comparing the STM topographic and spectroscopic images, we find that the atomic-scale protrusions on silicene show a dramatically low density of states over much of the energy range studied, which indicates that the electrons are more localized at these positions, as shown in Figure 2(d-f). Such a large difference in density of states cannot be attributed to a possible impurity-induced structural distortion at the TL Si atoms. Most likely, the bright protrusions are raised by oxygen adatoms, because they always appear after oxidation, but never for pure silicene layers. To account for this observation, we carried out i*n-situ* Raman spectroscopic measurement on samples exposed to oxygen as shown in Figure 2(g). The spectra show a clear broad shoulder at lower wavenumber (450-510 cm$^{-1}$) next to the silicene signature $E_{2g}$ peak. The position and broadness of this peak are associated with the Si-O bonds due to Si *sp*$^3$ hybridization.[25,26] By increasing the oxygen dose, the intensity of the shoulder peak is enhanced in all the silicene structures. Therefore, we conclude that these bright protrusions are oxygen adatoms adsorbed on the silicene surface.

In our STM measurements, the oxygen adatoms are identified on bridge sites, resulting in the configuration of double-atom-bonding overbridging O atoms ($O^d$). These oxygen adatoms overbridge neighboring Si atoms, leading to Si-O-Si bonds in silicene oxide. Figure 3(a-c) suggests that $O^d$ is a major configuration in partially oxidized $\sqrt{13}\times\sqrt{13}$, 4×4, and $2\sqrt{3}\times2\sqrt{3}$ silicene layers. Although $O^d$ is present in all silicene structures, the heights of oxygen adatoms residing on silicene layers are different, as demonstrated by the STM images in Figure 3(d). Oxygen adatoms on $\sqrt{13}\times\sqrt{13}$ and $2\sqrt{3}\times2\sqrt{3}$ silicene layers appear to be higher than those adsorbed on 4×4 silicene by about 1 Å. The distances between nearest neighboring TL Si atoms are 5.46 Å and 3.67 Å for $\sqrt{13}\times\sqrt{13}$ and $2\sqrt{3}\times2\sqrt{3}$ silicene, respectively. These distances are longer than twice the typical Si-O bond lengths in bulk $SiO_2$ (varies from 1.58 Å to 1.62 Å). Therefore, both TL and BL Si atoms are involved in silicon-oxygen bonds as Si(BL)-$O^d$-Si(TL). Oxygen adatoms prefer to reside beside TL Si rather than BL Si, as shown in the STM results. By contrast, the distance between nearest neighboring TL Si atoms in 4×4 silicene is 2.51 Å, indicating different buckling from the other two superstructures. DFT calculations indicate that TL Si atoms in 4×4 silicene can decrease in height, forming BL Si atoms under oxidation, in order to minimize total energy. Therefore, Si(BL)-$O^d$-Si(BL) is also a possible configuration for overbridging oxygen adatoms. As a result, the height of oxygen adatoms on 4×4 silicene is the lowest among the three buckled superstructures. It should be noted that $O^d$ is an energetically favoured configuration for oxygen adatoms on all the three silicene structures, even when the oxygen dose is increased up to 60 L.

The above conjecture on oxygen adsorption configurations is further verified by the DFT calculations, in which oxygen adatoms prefer to adsorb on the bridge sites of silicene for all the three configurations upon relaxation. The equilibrium structures and adsorption energy for individual oxygen adatoms on silicene monolayers in the three typical superstructures are shown in Figure 3(e-g). The adsorption energy, $E_{\text{ads}}$, for an oxygen adatom on silicene is defined as

$$E_{\text{ads}} = E_{\text{tot}} - E_{\text{Si-Ag}} - \frac{1}{2} E_{\text{O2}} \qquad (1)$$

where $E_{\text{tot}}$ is the total energy of entire system of Ag(111)-supported silicene with an oxygen adatom; $E_{\text{Si-Ag}}$ is the total energy of the silicene superstructure on Ag(111); $E_{\text{O2}}$ is the energy of an oxygen molecule in gas phase. By definition, negative adsorption energy means that the oxidation of silicene is exothermic. The computed adsorption energy of about −3 eV indicates that silicene sheets in those three superstructures can be easily oxidized. Among them, √13×√13 silicene seems to be most easily oxidized due to the largest amplitude of adsorption energy. The distance between the oxygen adatom and the Ag(111) surface can be measured by a height parameter $d$ [see Figure 3(e-g)], which is taken the average height of all Ag atoms in the first layer of the slab model as reference. Among the three superstructures considered, the 4×4 has the lowest height of 3.33 Å, which is about 0.5 Å lower than those of the √13×√13 ($d$ = 3.84 Å) and 2√3×2√3 ($d$ = 3.88 Å) structures. The relative height difference among the three different structures is in excellent accordance with experimental observations by STM, as shown in Figure 3(d). It should be noted that the DFT calculations are only able to qualitatively explain the experimental observations. For each superstructure, we only considered one oxygen adatom per unt cell within periodic boundary condition; which may differ from the realistic situation in both concentration and spatial distribution of the oxygen atoms. Nevertheless, the difference between various oxidized silicene superstructures revealed from DFT simulations would be still valid.

It has been predicted that the band structure of silicene can be tailored into various types, including semimetals, semiconductors, and insulators, by chemical functionalizations such as oxidation.[18,20] The STM and STS results on partially oxidized silicene layers on Ag(111) are shown in Figure 4. It displays a series of spectra taken along lines cut across the oxygen adatoms on the three silicene superstructures. The magnitude of the gap shows significant variation corresponding to the different superstructures. Another common characteristic is that the gap is larger at oxygen adatom sites and becomes smaller in the locations away from the absorption sites. Despite that, the gap still exists in the lateral distance of 3 nm around oxygen adatoms, which indicates that the oxygen adatoms could affect the electronic structure of silicene in a large area. Since the average distances between neighbouring oxygen adatoms on silicene in each structure are less than 3 nm, this suggests that the gap is opened over the whole silicene surface due to adsorption of oxygen adatoms even with low oxygen dose of 20 L. It should be noted that the oxygen adatoms do not show a ordered structure, which might lead to variations of gap value at different sites on oxidized silicene surface. In 4×4 silicene, the gap varies from 0.18 to 0.30 eV under oxygen dose of 20 L. The most typical gap value is about 0.18 eV, as shown in Figure 4(b). While √13×√13 and 2√3×2√3 structures show band-gap values of 0.11 eV - 0.14 eV and 0.15 eV - 0.18 eV, respectively, as shown in Figure 4(a) and (c). These values of band gaps are in qualitative agreement with DFT calculations shown in supporting information. Because pure silicene in each structure exhibits a characteristic semimetal zero gap, this clearly demonstrates that there is a band-gap opening associated with the amount of oxygen adatoms. The band gap is increased with increasing oxygen dose. The gap of oxidized silicene is homogeneous when oxygen dose is greater than 30 L (see Supporting Information). The band gaps are 0.18 eV, 0.9 eV, and 0.22 eV for the √13×√13, 4×4, and 2√3×2√3 structures under oxidation with an oxygen dose of 60 L, respectively. These values are well below the width of the semiconducting band gap in bulk silicon. While the gap opens homogeneously for oxidized silicene (oxygen dose >

30L), small differences in occupied and unoccupied states can be observed, which are most likely due to the inhomogeneous local density of states induced by disordered oxygen adatoms. According to a previous DFT simulations,[20] the conduction band of partially oxidized silicene is mainly contributed by the Si *p*-orbital and O *p*-orbital, and the valence band originates from the O *p*-orbital. The width of the band gap is predominantly influenced by the adsorption sites of oxygen adatoms. Since the valence band of silicene oxide mainly originates from the *p*-orbital of O, the dangling bonds of TL Si in oxidized 4×4 silicene are fully saturated by oxygen adatoms, which results in the largest gap in the oxidized silicene among the three structures. Unpaired electrons in oxidized √13×√13 and 2√3×2√3 silicene layers, however, contribute a narrow gap under low oxygen doses. By varying the oxygen dose, we found that the band gaps are indeed tunable and dominated by the coverage of oxygen adatoms. In contrast to graphene,[21] it is found that oxygen adatoms prefer to be accommodated at the surface of silicene rather than the edge, which is most likely because the dangling bonds on edge Si atoms are passivated by the Ag(111) surface.

Recent studies have claimed that silicene can only be oxidized at a high oxygen dose of 1000 L.[18] It is still unclear, however, how oxygen adatoms associate with silicene layers, especially with respect to their structure before the start of oxidation, which is critical for further chemical functionalization. Figure 5 shows STM results on silicene exposed to different oxygen doses, *i.e*. 10 L, 60 L, and 600 L, respectively. It is found that the surface of silicene in the 2√3×2√3 structure is fully covered by oxygen adatoms at the oxygen dose of 60 L [Figure 5(b)] and exhibits an amorphous-like disordered feature instead of a distinct structure. The insets in Figure 5 show the corresponding fast Fourier transform (FFT) patterns for each sample. Interestingly, clear FFT patterns with bright symmetric spots were observed for 2√3×2√3 silicene exposed to 10 L and 60 L $O_2$, indicating that partially oxidized silicene retains the hexagonal honeycomb structure. Nevertheless, the FFT pattern shows a typical amorphous feature, indicating the full oxidation of 2√3×2√3 silicene phase when the oxygen dose was increased to 600 L. Moreover, it is worth noting that some areas of bare Ag(111) substrate were observed in the fully oxidized silicene, which has not been reported in either experimental or theoretical works before. The evolution of such "silicene-free" areas can be explained by comparing the binding energies of AgO and $SiO_2$. The binding energy between the epitaxial silicene layer and the Ag(111) surface is about 0.7 eV,[22] which is much smaller than the binding energy for Si-O (between 4.0 and 12.0 eV).[23] The oxygen thus tends to bond firstly with the Si atoms in the silicene instead of the Ag atoms in the substrate. Moreover, the energy required for the oxygen adsorption on Ag(111) is much higher than on the Si surface with dangling bonds, and therefore, bare Ag(111) surface rather than silver oxide appears in the fully oxidized silicene sample.[22,28] Due to the characteristic $sp^3$ hybridization of Si, energetically stable Si-O-Si bonds would be expected when silicene is exposed to a high oxygen dose (600 L).

**CONCLUSION**

In summary, an electronic band gap in monolayer silicene on an Ag surface was induced by oxidation, which was verified by STM and *in-situ* Raman spectroscopy studies. $O^d$ is the most energetically favoured configuration for the adsorption of oxygen adatoms on the surfaces of √13×√13, 2√3×2√3, and 4×4 silicene. The different buckled structures lead to different heights of oxygen adatoms on the silicene. The band gap can be modulated from semimetallic to semiconducting type, which can very well overcome the zero-gap disadvantage of silicene. In fully oxidized silicene, the buckled silicene structure vanishes, with subsequent crumpling of the sample and exposure of bare Ag(111) surface areas.

**METHODS**

**Materials.** All samples used in this work were synthesized *in-situ* in a preparation chamber of a low-temperature ultra-high-vacuum (UHV) scanning tunnelling microscopy/scanning near-field optical microscopy system (LT-STM-SNOM) (SNOM1400, Unisoku Co.). A clean Ag(111) substrate was prepared by argon ion sputtering and subsequent annealing at 820 K for several cycles. The silicene monolayers were fabricated by the evaporation of silicon from a heated silicon wafer. The deposition flux was 0.08 monolayers per minute (ML/min). The temperature of the Ag(111) substrate was 450 K, 500 K, and 550 K for the formation of $\sqrt{13}\times\sqrt{13}$, 4×4, and $2\sqrt{3}\times2\sqrt{3}$ phases, respectively. Silicene oxide samples were prepared by an *in-situ* oxidation of silicene monolayers with a varying $O_2$ dose during oxidation in the chamber. The Langmuir (L) is used as the unit of exposure of $O_2$, *i.e.* 1 L is an exposure of $10^{-6}$ torr $O_2$ in one second.

**Characterization.** The STM and Raman spectroscopy measurements were carried out in ultra-high vacuum (UHV, $< 8\times 10^{-11}$ torr) at 77 K. Scanning tunneling spectroscopy (STS) differential conductance ($dI/dV$) (where *I* is current and *V* is voltage) was performed with lock-in detection by applying a small modulation of 20 mV to the tunnel voltage at 973 Hz. The differential conductance maps were obtained by recording an STS spectrum at each spatial pixel during STM topographic measurements. Before the STS measurements, the Pt/Ir tip was calibrated on a silver surface. The Raman spectra were acquired using a laser excitation of 532 nm (2.33 eV) delivered through a single-mode optical fibre.

**DFT calculation.** *Ab initio* calculations were performed using density functional theory (DFT) and the plane wave basis, as implemented in the Vienna *Ab initio* Simulation Package (VASP).[29] The electron-ion interactions were represented by projector augmented wave (PAW) potentials.[30]

The generalized gradient approximation (GGA) with the Perdew-Burke-Ernzerhof (PBE) functional was adopted to describe the exchange-correlation interaction.[31] A kinetic energy cut-off of 400 eV for the plane-wave basis and a convergence criterion of $10^{-4}$ eV for the total energies were carefully tested and adopted in all calculations. The structures for the superstructures of silicene on the Ag(111) surface were derived from a previous simulation by Gao and Zhao.[4]

**Corresponding Author**

* To whom correspondence should be addressed: ydu@uow.edu.au

**Acknowledgements**

This work was supported by the Australian Research Council (ARC) through a Discovery Project (DP 140102581) and Linkage, Infrastructure, Equipment and Facilities (LIEF) grants (LE100100081 and LE110100099), and by the National Natural Science Foundation of China (11134005). Y. Du would like to acknowledge support by the University of Wollongong through the Vice Chancellor's Postdoctoral Research Fellowship Scheme and a University Research Council (URC) Small Grant. We are grateful to Prof. Weichang Hao at Beihang University for valuable comments.

**Supporting Information**. The detailed methods, characterizations, oxidation process of silicene layers, in-situ Raman spectroscopy results on different silicene structures, and identification of sites of oxygen adatoms on silicene layers are included in the Supporting Information. This material is available free of charge *via* the Internet at http://pubs.acs.org.

**FIGURES**

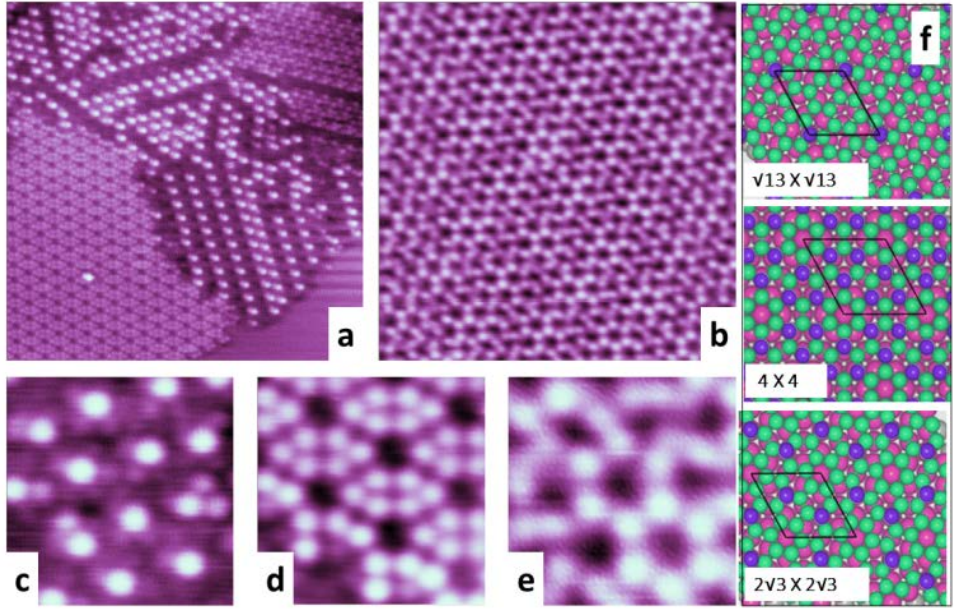

**Figure 1.** Topographic images of silicene monolayers grown on Ag(111): (a) STM topographic image of two major coexisting phases of silicene, √13×√13 and 4×4, which occupy different domains (scanning area 16 nm × 16 nm, $V_{bias}$ = -0.5 V, $I$ = 4 nA). (b) STM topographic image of silicene 2√3×2√3 phase (scanning area 10 nm × 10 nm, $V_{bias}$= -0.8 V, $I$ = 4 nA). (c)-(e) High-resolution STM images of √13×√13, 4×4, and 2√3×2√3 phases, respectively (scanning area 2 nm × 2 nm, $V_{bias}$ = -0.02 V, $I$ = 5 nA). (f) Illustrations of various phases of silicene monolayers on Ag(111), in which dark balls represent Si atoms and light balls represent underlayer Ag atoms in Ag(111). Purple and green atoms represent top-layer and bottom-layer Si atoms, respectively, in buckled-structured silicene. Pink atoms represent Ag atoms in the substrate.

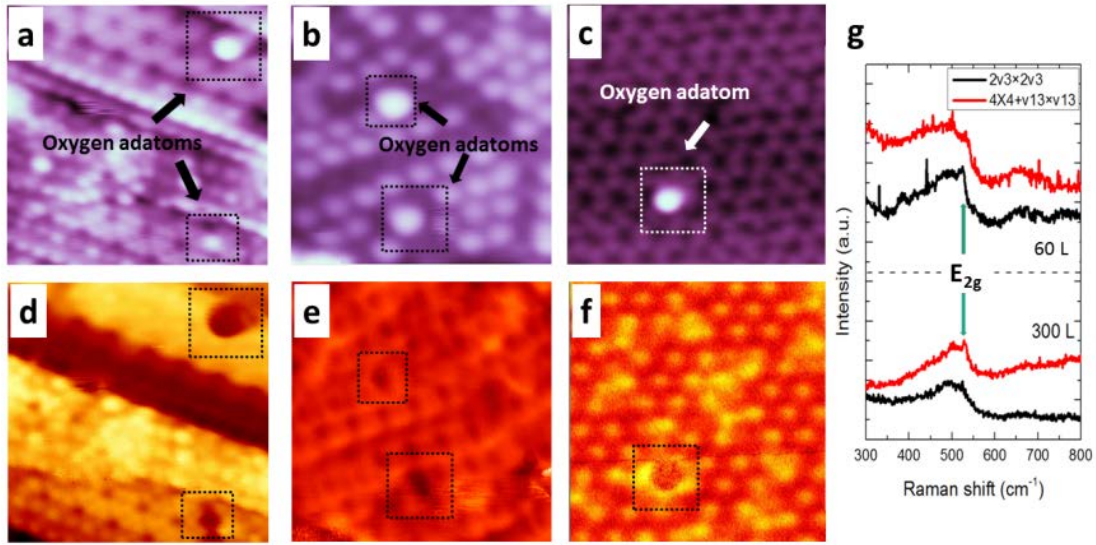

**Figure 2.** STM images of oxygen adatoms on silicene in (a) $\sqrt{13}\times\sqrt{13}$, (b) 4×4, and (c) $2\sqrt{3}\times2\sqrt{3}$ phases grown on Ag(111) substrate (scanning area 4 nm × 4 nm, $V_{bias}$ = -0.2 V, $I$ = 4 nA). The bright protrusions are attributed to oxygen adatoms in each STM image, which are indicated by the arrows. (d)-(f) Corresponding STS mappings to STM images (a)-(c), respectively. (g) *In-situ* Raman spectra for silicene oxidized under different oxygen doses. An obvious broad shoulder at lower wavenumber to the $E_{2g}$ peak indicates the formation of Si-O bonds.

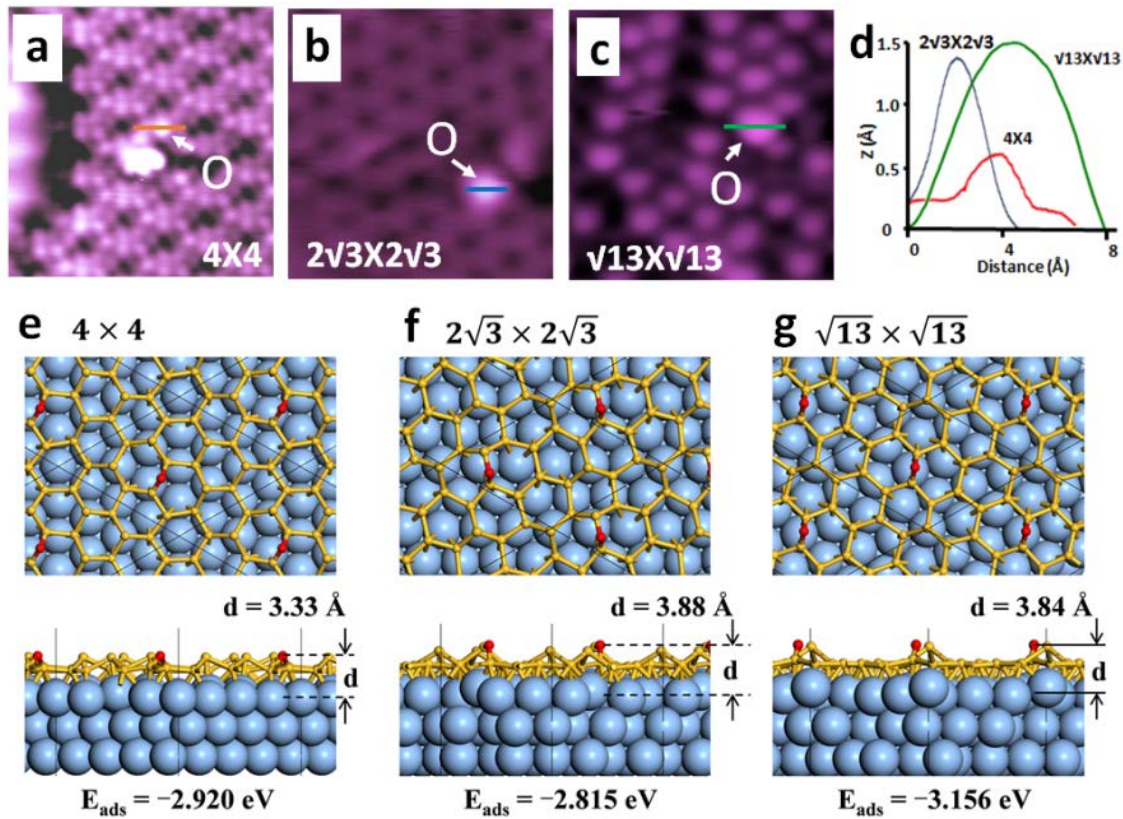

**Figure 3.** STM and STS images of oxidized silicene in (a) 4×4, (b) 2√3×2√3, and (c) √13×√13 structures (scanning area 4 nm×4 nm, $V_{bias}$ = -0.2 V, $I$ = 4 nA). The oxygen adatoms prefer to reside on TL Si atoms in the initial oxidation. (d) Line profiles of oxygen adatoms on silicene corresponding to the lines in the STM images in (a), (b), and (c), respectively. (e)-(g) DFT simulations (top and side views) of atomic structures for oxygen adatoms on Ag(111) supported silicene monolayers in different superstructures: (e) 4×4, (f) 2√3×2√3, (g) √13×√13. The black rhombuses in the top views represent the unit cell. Red: oxygen; yellow: silicon; blue: silver.

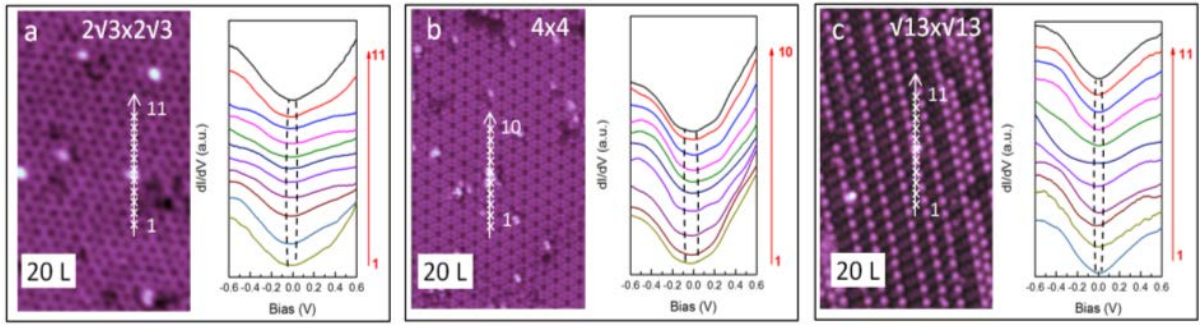

**Figure 4.** Spatial evolution of the electronic states measured on (a) 2√3×2√3, (b) 4×4, and (c) √13×√13 silicene exposed under oxygen dose of 20 L. Tunneling spectra ($dI/dV$ curves) were obtained along a line denoted by the arrows in the corresponding STM topographic images on the right. The dashed lines in each STS result illustrate the value of band gap. STM images were obtained at $V_{bias} = -0.8$ V, $I = 0.6$ nA. The oxygen adatoms appear as bright protrusions on the silicene layers.

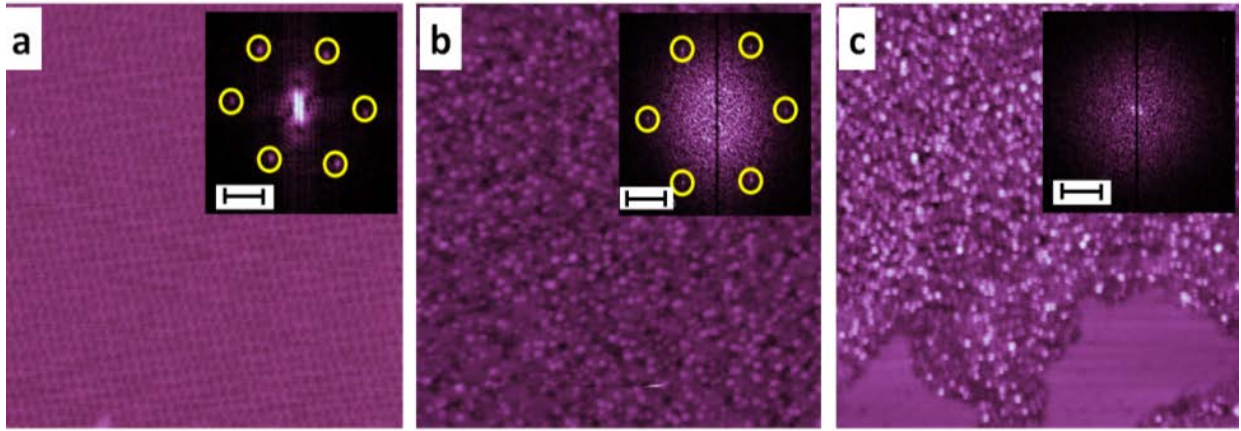

**Figure 5.** STM images of 2√3×2√3 silicene oxidized under (a) 10 L $O_2$, (b) 60 L $O_2$, and (c) 600 L $O_2$. The insets show the corresponding FFT patterns. Bare Ag(111) surface can be seen in the bottom area of (c). The scale bar in each inset represents 1/nm.

**ToC/Abstract graphic**

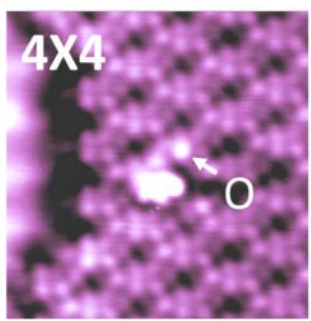 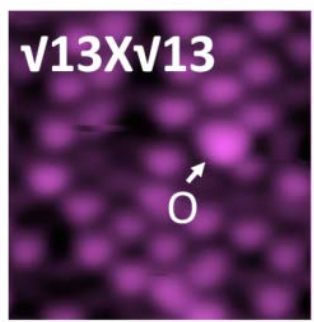 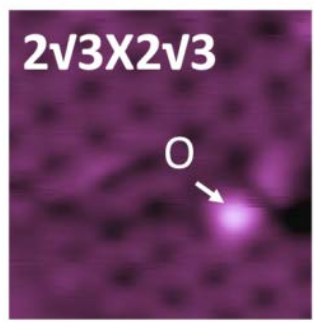

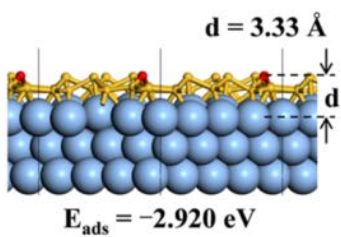 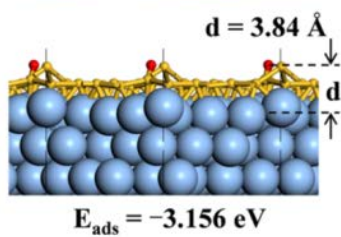 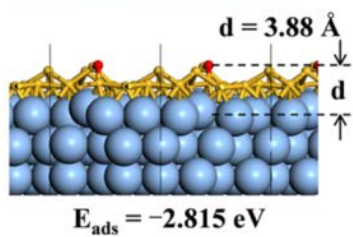